\documentstyle[11pt,newpasp,twoside,epsf]{article}
\def\edcomment#1{\iffalse\marginpar{\raggedright\sl#1\/}\else\relax\fi}
\reversemarginpar

\begin{document}
\title{Galaxy threshing and the origin of intracluster stellar objects}
\author{Kenji Bekki, Warrick J. Couch}
\affil{School of physics, University of New South wales, Sydney, NSW, 2052, Australia
}
\author{Michael J. Drinkwater}
\affil{Department of Physics, University of Queensland, Queensland 4072, Australia}
\author{Yasuhiro Shioya}
\affil{Astronomical Institute, Tohoku University, Sendai, 980-8578, Japan}

\begin{abstract}

We numerically investigate dynamical evolution of non-nucleated
dwarf elliptical/spiral  galaxies (dE) and nucleated ones (dE,Ns) in
clusters of galaxies in order to understand the origin 
of intracluster stellar objects, such as intracluster stars (ICSs),
GCs (ICGCs), and ``ultra-compact dwarf'' (UCDs)
recently discovered by  all-object spectroscopic survey
centred on the Fornax cluster of galaxies.
We find that the outer stellar components of a nucleated dwarf 
are removed by the strong tidal field of the cluster,  
whereas the nucleus manages to survive as a result of 
its initially compact nature. The developed naked nucleus is found 
to have physical properties (e.g., size and mass) similar to those 
observed for UCDs. We also find that 
the UCD formation processes  does depend on the radial density profile of 
the dark halo in the sense that UCDs are less likely to be formed from 
dwarfs embedded in dark matter halos with central `cuspy' density 
profiles. Our simulations also suggest that very massive and compact 
stellar systems can be rapidly and efficiently formed in the central 
regions of dwarfs through the merging of smaller GCs. 
GCs initially in the outer part of dE  and dE,Ns are found to be stripped 
to form ICGCs.  
\end{abstract}

\section{UCD as an intracluster stellar object}

A new type of sub-luminous and extremely compact ``dwarf galaxy'' has
recently been discovered in an ``all-object''  spectroscopic survey
centred on the Fornax cluster of galaxies (Drinkwater et al. 2000).
While objects with this type of {\it morphology} have been observed 
before -- the bright compact objects discovered by Hilker et 
al. 1999 -- and the very luminous globular clusters around cD galaxies 
(Harris, Pritchet, \& McClure 1995) -- in this particular case they
have been found to be members of the Fornax cluster, 
have intrinsic sizes of only $\sim$ 100\,pc, and have absolute
$B-$band magnitudes ranging from $-13$ to $-11$\,mag. Hence Drinkwater et 
al. have named them ``ultra-compact dwarf'' (UCD) galaxies.
Importantly, the luminosities of UCDs are intermediate between those of 
globular clusters and small dwarf galaxies and  are similar to those of the 
bright end of the luminosity function of the nuclei of nucleated dwarf 
ellipticals. 

Radial distribution, orbital velocity dispersion, and metallicity distribution
of UCDs are suggested to provide valuable information on the difference
in formation histories between UCDs, ICGCs, and ICSs
(Bekki et al. 2003a). 
The ``galaxy threshing'' scenario (Bekki et al. 2001) has predicted that
only luminous dE,Ns with highly eccentric orbits and small pericenter distance
from the center of a cluster can become UCDs after the outer dwarf 
envelopes are completed stripped by the cluster tidal field. 
ICGCs and ICSs have been demonstrated to form
via tidal stripping of GCs and stars from cluster member galaxies
(Bekki et al. 2003b).  
Here we reinvestigate
the formation of UCDs/ICGCs/ICSs  by using numerical simulations with larger number of
particles (up to $N \sim 10^6$) to understand 
(1) how the formation histories of UCDs depend on the structure of dE,Ns
(in particular, the central density of their dark matter halos, i.e., cores vs cusp), 
(2) the dynamical evolution of GCs
in dEs orbiting the Fornax cluster ,  (3) whether these GCs can become UCDs
in the center of dEs via merging of GCs.
The details of the models for the Fornax cluster are given in Bekki et al. (2003a)
and thus we briefly summarize the results here.

\section{Galaxy threshing and UCD/ICS formation}

Figures 1 and 2 summarize the dynamical evolution of 
the dE,N model with $M_{\rm B}$ = $-16$ mag,
the NFW dark matter halo, and the nuclear mass fraction of 0.05
(referred to as the fiducial model, FO1).
As the dE,N approaches the pericenter of its orbit,
the strong global tidal field of the Fornax cluster 
stretches the envelope of the dE'N along the direction of the dwarf's orbit 
and consequently tidally strips the stars of the envelope ($T = 1.13$\,Gyr). 
The dark matter halo, which is more widely distributed than the envelope 
due to its larger core radius, is also efficiently removed from the dE,N 
during the pericenter passage. Since the envelope (and the dark matter halo) 
loses a significant fraction of its mass during the passage of the 
pericenter, the envelope becomes more susceptible to the tidal effects of 
the Fornax cluster after the pericenter passage. Therefore, each subsequent
time the dwarf approaches the pericenter, it loses an increasingly larger
fraction of its stellar envelope through tidal stripping (compare, for
example, the $T = 2.26$ and $T = 2.83$\,Gyr time points). 
Consequently, both the envelope and the dark matter halo become
smaller,  less massive,  and more diffuse after five passages of 
the pericenter  ($T = 3.34$\,Gyr).

\begin{figure}
\caption{
Morphological evolution of the stellar components (the stellar envelope
and the nucleus) of the dE,N projected onto the $x$-$y$ plane for the 
fiducial model (FO1). The time $T$ (in units of Gyr) indicated in the 
upper left corner of each frame represents the time that has elapsed since
the simulation starts. Each frame is 500\,kpc on a side in the upper six  
panels and 17.5\,kpc on a side in the lower panels.
The orbital evolution of the dE,N at 4.5\,Gyr is 
indicated by the solid line in the upper left panel in the 
upper six panels. 
The location of the nucleus of the dwarf is indicated
by crosses for $T$ = 1.13, 2.26, 2.83, and 3.34 Gyr.
The scale radius of the adopted NFW model for the dark matter halo
distribution of the Fornax cluster mass profile
is indicated by a dotted circle in each of the upper six panels.
}
\end{figure}

\begin{figure}
\plotone{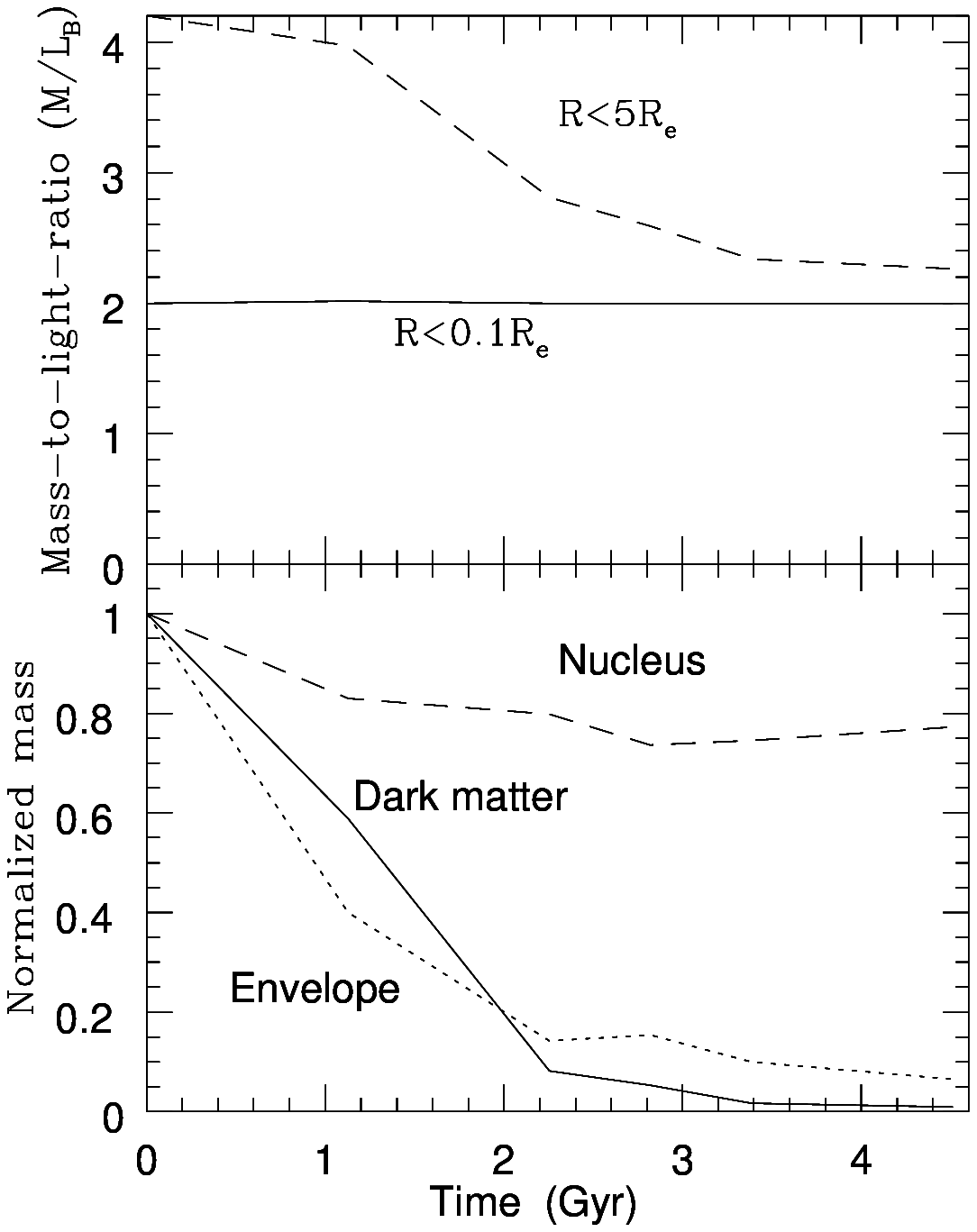}
\caption{
Time evolution of the $B$--band mass-to-light-ratio ($M/L_{\rm B}$)
({\it upper} panel) and that of the total mass normalized to the initial 
mass for each collisionless component ({\it lower} panel) in the fiducial 
model (FO1). In the {\it upper} panel, $M/L_{\rm B}$ estimated for $R$ $<$ 
$0.1R_{\rm e}$ (where $R$ and  $R_{\rm e}$ are the distance from the center 
of the dE,N and the initial effective radius of the dE,N, respectively)
and for $5R_{\rm e}$, are shown by the {\it solid} and  {\it dotted} lines,  
respectively. In the {\it lower} panel, the total mass within $R$ $<$ 
$5R_{\rm e}$, $R$ $<$ $R_{\rm e}$, and $R$ $<$ $0.1R_{\rm e}$
for the dark matter halo, the stellar envelope, and the nucleus is
shown by the {\it solid, dotted} and {\it dashed} lines, respectively.
The mass-to-light ratio, $M/L_{\rm B}$, decreases 
dramatically from 4.2 to 2.3  for $r < 5r_{\rm e}$. This result clearly 
explains why the UCDs are observed to have mass-to-light ratios that are
much smaller ($M/L_{\rm B} = 2$--4; D03) than those observed for dE,Ns ($\sim$ 
10) for some of the Local group dwarf: 
galaxy threshing is the most efficient in the outer regions of a dE,N 
where the dark matter halo dominates gravitationally. About 95\% of the 
envelope initially within $r_{\rm e}$, and 20\% of the nucleus initially 
within $0.1r_{\rm e}$ are removed from the dE,N. 
}
\end{figure}

The stripped stars form a long tidal stream of ICSs along the ``rosette'' orbit 
within the orbital plane ($T = 1.13$\,Gyr) and the ICSs can have metallicity
of [Fe/H] $\sim$ $-0.76$ for $B-V$ = 1 for the dE,N.  
The central nucleus, on the other hand, is just weakly influenced by the
tidal force as a result of its compact configuration. Because of its 
strongly self-gravitating nature, the nucleus loses only a small amount 
($\sim 20$\%) of its mass and thus maintains its compact morphology during 
its tidal interaction with the Fornax cluster. As a result, a very 
compact stellar system with a negligible amount of dark matter
is formed from the dE,N by $T = 4.5$\,Gyr. The total mass and 
size of the remnant are $\sim$ 3.8 $\times$ $10^7$ $M_{\odot}$ and
$\sim$ 100 pc (five times the core or scale radius), 
consistent with the observed properties of a UCD. Hence
this new study, based on fully self-consistent numerical models of dE,Ns, 
confirms the earlier results of Bekki et  al. (2001) based on a more
simplistic model.

\section{GC evolution in cluster dEs: UCD or ICGC formation ?}

Our simulations on GC merging in dEs/dIs
have found the following  interesting results: (1) Galactic nuclei or UCDs formed
by $10-20$ GC merging
form a scaling relation that is different both from GCs and Es (Figure 3),
(2) The shapes of the developed nuclei or UCDs depends on the host dwarf's
luminosity in the sense that brighter dEs have more spherical nuclei/UCDs,
(3) The brighter dEs are more likely to be transformed into dE,Ns,
(4) The GCs in the outer part of dEs ($R$ $>$ $0.5R_{\rm e}$) are 
likely to be stripped to form ICGCs within a few Gyr (thus only
central GCs can be merged into the nuclei/UCDs),  
and (5) Structure and
kinematical properties of these ICGC systems depend on the orbital
properties of their host dwarfs.

The present set of numerical simulations suggest that morphological evolution
of dEs/dIs with GCs can be different between these 
after they enter into a cluster environment:
Some of them can be transformed into dE,Ns via GC merging in their central regions,
if initial distributions of their GCs are more centrally concentrated.
Such newly formed dE,Ns can be furthermore transformed into UCDs
due to the strong cluster tidal field, if
it has lower dark matter density, highly eccentric orbits, and small pericenter
distance.
If dE/dIs with GCs have  more diffusely distributed GC systems,
their GCs are tidally stripped to form ICGCs with the host dwarfs' $S_{\rm N}$
decreasing significantly.
Some fraction of the observed 
ICSs can originate from the disrupted dEs and dE,Ns (by galaxy threshing)
in a cluster so that
the spatial distribution, kinematics, and  metallicity distribution
of these ICSs reflect the orbital evolution and the disruption process
of these dwarfs in the cluster.
Thus we need to understand first the 
dynamical evolution of low luminosity cluster dwarfs,
which are numerous and more susceptible to  disruption
by cluster global tidal field (and thus one of major sources of
intracluster stellar objects)
in order to understand
the origin of UCDs, ICGCs, and ICSs.

\begin{figure}
\plotone{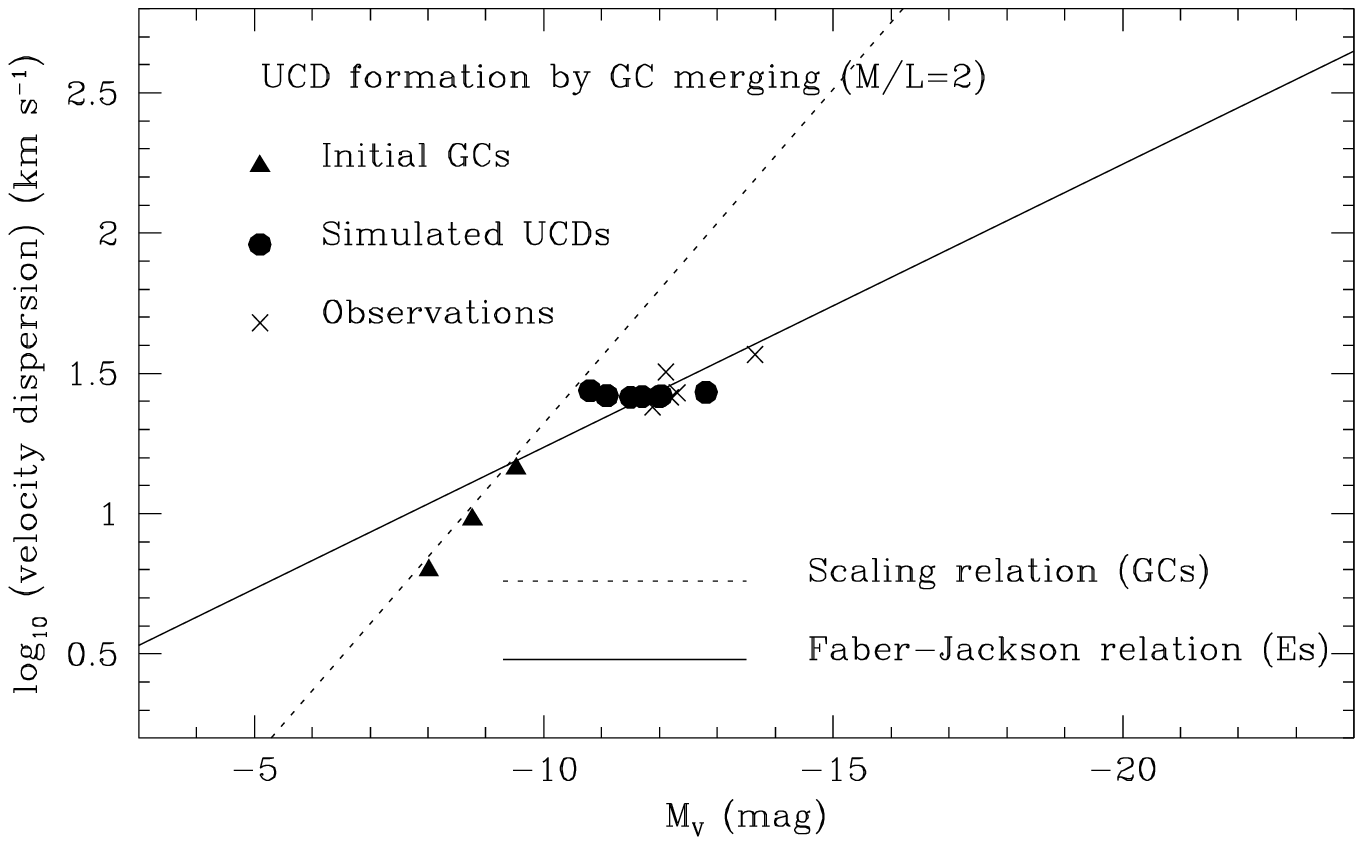}
\caption{
Distribution of UCDs formed from GC merging 
in dEs/dIs on the 
[central velocity dispersion, $M_{\rm V}$]--plane.
Six UCDs formed in the central regions of dEs via GC merging
are represented by the {\it filled circles}. The central velocity dispersion is
given in units of km s$^{-1}$ and plotted on a ${\rm log}_{10}$ scale. 
For comparison, the observed UCDs ({\it crosses}) and the original GCs 
({\it filled triangles}) are also plotted. The {\it solid} and the {\it dotted}
lines represent the scaling relations for GCs (Djorgovski 1993) and elliptical galaxies
(Faber \& Jackson  1976), respectively. The model and the method for
deriving $M_{\rm V}$ and the central velocity dispersion of the
UCDs formed from GC merging are given in Bekki et al. (2003a).
The total number of GCs ($N_{\rm GC}$) is 19 for the dE model with $M_{\rm B}=-16$ mag and
$S_{\rm N}$ = 5 and these $N_{\rm GC}$ and $S_{\rm N}$ are considered to
be free parameters in this numerical study. The initial GC mass ranges from
$10^5 M_{\odot}$ to $2 \times 10^6 M_{\odot}$, and  $N_{\rm GC}$ and $S_{\rm N}$ 
are changed according to the adopted
mass of each GC. Note that the developed massive star clusters via GC merging,
which can be identified as ``nuclei'' or ``UCDs'' in the center of dEs,
have a scaling relation different from those of Es and GCs. This result
implies that the observed location of UCDs in this plane can be understood
in term of GC merging in the dE's central regions. 
The shapes of galactic nuclei (UCDs) depend on whether the host dwarfs
have elliptical shapes (dE) or disky ones (dI).
Some nuclei/UCDs  have rotation and flattened shapes, in particular, 
for dIs with smaller number of GCs.
The more detailed discussion on this result and on the relation of nuclei/UCDs to
$\omega$ Cen and G1 are given in Bekki et al. (2003a).
}
\end{figure}

\end{document}